\documentclass[prb,aps,twocolumn,superscriptaddress]{revtex4-2}
\usepackage{amsmath}
\usepackage{subfigure}
\usepackage{color}
\usepackage{bbm}
\usepackage{amssymb}
\usepackage{epsfig}
\usepackage{multirow}
\usepackage{amsbsy}
\usepackage{array}
\usepackage{diagbox}
\usepackage{bm}
\usepackage{extarrows}
\usepackage{graphicx}
\usepackage{appendix}
\usepackage{txfonts}
\usepackage{lipsum} 
\usepackage{tablefootnote}

\usepackage[colorlinks=true,linkcolor=blue,citecolor=blue,urlcolor=blue,bookmarks=false]{hyperref}

\usepackage{makecell}
\usepackage[percent]{overpic}
\begin{document}
\title{A variational Monte Carlo approach to the SU(4) spin-orbital model on the triangular lattice}

\author{Chun Zhang}
\affiliation{Institute of Physics, Chinese Academy of Sciences, Beijing 100190, China}

\author{Hui-Ke Jin}
\affiliation{Technical University of Munich, TUM School of Natural Sciences, Physics Department, 85748 Garching, Germany}

\author{Yi Zhou}
\email{yizhou@iphy.ac.cn}
\affiliation{Institute of Physics, Chinese Academy of Sciences, Beijing 100190, China}
\affiliation{Kavli Institute for Theoretical Sciences, University of Chinese Academy of Sciences, Beijing 100190, China}
\affiliation{CAS Center for Excellence in Topological Quantum Computation, University of Chinese Academy of Sciences, Beijing 100190, China}

\date{\today}

\begin{abstract}
Previous investigations have suggested that the simplest spin-orbital model on the simplest frustrated lattice can host a nematic quantum spin-orbital liquid state. Namely, the orbital degeneracy of the SU(4) Kugel-Khomskii (KK) model tends to enhance quantum fluctuations and stabilize a quantum spin-orbital liquid exhibiting stripy features on the triangular lattice, as revealed by the state-of-the-art method of the density matrix renormalization group boosted by Gutzwiller projected wave functions. In this work, using the variational quantum Monte Carlo method, we have studied several spin-orbital liquid states, including a uniform $\pi$ flux state, three stripy states, and a plaquette state, on the $L\times{}L$ torus up to $L=24$. It turns out that one of these stripy states, called the ``stripe-II'' state, is energetically favored. This ground state breaks the $C_6$ symmetry of the lattice, resulting in a reduced $C_2$ symmetry and doubled unit cells, while preserving the SU(4) spin-orbital rotation symmetry. Such a nematic quantum spin-orbital liquid state can be characterized by a parton Fermi surface (FS) consisting of open orbits in the Brillouin zone, in contrast to the circular FS of the uniform $\pi$-flux state.
\end{abstract}

\maketitle

\section{Introduction}

The search for quantum spin liquids (QSLs) is one of the central issues in modern condensed matter physics~\cite{Anderson1973,Lee08,Balents10,Zhou2017,Savary2016,Knolle2019,Broholm2020}. In the 1970s, P. W. Anderson proposed a disordered ground state for the spin-1/2 antiferromagnetic (AFM) Heisenberg model on the triangular lattice~\cite{Anderson1973}, namely the simplest spin model on the simplest geometrically frustrated lattice, which revealed the age of QSLs and resonating valence-bond (RVB) states.
However, after extensive studies, the academic community has reached a consensus that the spin-1/2 AFM Heisenberg model on the triangular lattice hosts a 120$^\circ$ magnetically ordered ground state, contrary to Anderson's original proposal~\cite{Huse1988,Jolicoeur1989,White2007}. It means that the smallest spin quantum $S=1/2$, which gives rise to the strongest quantum spin fluctuations, is still insufficient to completely suppress the classical magnetic order on the triangular lattice, if only the nearest neighbor (NN) Heisenberg spin interaction $\vec{S}_{i}\cdot\vec{S}_{j}$ is considered.
Therefore, one has to turn to other ways of enhancing quantum fluctuations to realize a QSL ground state on the simplest geometrically frustrated lattice.

In addition to (i) the small spin quanta $S$ and (ii) geometric frustration, there are several other ways to enhance quantum spin fluctuations~\cite{Jin2022CPL}: (iii) competing spin interactions involving NN anisotropic spin couplings (e.g., the anisotropic Ising couplings in the Kitaev honeycomb model~\cite{Kitaev06}) and isotropic (or anisotropic) spin couplings on longer bonds; (iv) charge fluctuations near a Mott transition leading to multiple-spin interactions; and (v) additional degeneracy due to orbital degrees of freedom.
Thus, QSL states can be achieved by stronger geometric frustration~\cite{Yan2011, Depenbrock2012, Liao2017, He2017} or by increasing competition among mutual interactions~\cite{Block2011, Mishmash2013, Zhu2015, Hu2015, Wietek2017, Gong2017, Saadatmand2017, Hu2019}. 
In particular, orbital degeneracy [i.e. the path (v) mentioned above] can significantly enhance quantum fluctuations~\cite{Feiner1997}. For instance, in some transition metal oxides, spin and orbital degrees of freedom may play a symmetric role, resulting in an enlarged SU(4) rather than the SU(2)$\times$SU(2) spin-orbital rotational symmetry. The SU(4) Kugel-Khomskii (KK) model is the minimal model used to describe such spin-orbital materials~\cite{Kugel1982, Li1998, Pati1998, Khaliullin2000, Tokura2020, Reynaud2001, Yamada2018,Yamada2021}, in which the SU(4) symmetry can amplify quantum fluctuations and potentially stabilize a spin-orbital liquid ground state.

Much theoretical effort has been devoted to the study of quantum phases in SU(4) quantum magnets. As an example, the one-dimensional SU(4) KK model is found to be integrable and to possess gapless excitations~\cite{Sutherland1975,Li1999}. These gapless excitations are well characterized by the low-energy effective theory, i.e. the SU(4)$_1$ Wess-Zumino-Witten (WZW) conformal field theory (CFT)~\cite{Affleck1986,Azaria1999,fuhringer2008}. The ground state of the SU(4) KK model on the two-leg ladder is also well known, which breaks the translational symmetry and forms an SU(4) singlet plaquette~\cite{Bossche2001,Weichselbaum2018}. This SU(4) singlet plaquette state can even be an exact ground state on the ladder, as long as the extra interactions have been properly chosen and added to the model Hamiltonian ~\cite{Chen2005}.  Unlike one-dimensional (1D) and ladder models, two-dimensional (2D) models are less clear, despite extensive research. For instance, several complementary methods indicated that the ground state of the SU(4) KK model on the honeycomb lattice is a Dirac-type spin-orbital liquid~\cite{Corboz2012,Andrade2019,Jin2022}. For the square lattice, several candidate ground states have been proposed for the SU(4) KK model, such as the plaquette ordered state~\cite{Li1998,Bossche2000}, the $\mathbf{Z}_2$ spin-orbital liquid~\cite{Wang2009}, and the dimerized and SU(4) symmetry-breaking state~\cite{Corboz2011}.

Especially, the ground state of the SU(4) KK model on the triangular lattice has been debated until very recently. In the early days, exact diagonalization (up to 16 sites) and variational approach (up to 64 sites) suggested an SU(4)-singlet plaquette liquid ground state on which simple types of long-range correlations are suppressed~\cite{Penc2003}.
Recently, the combination study of density matrix renormalization group (DMRG) calculation on cylinders (with a circumference up to $L_y=4$) and field theory analysis proposed a gapless liquid state with an emergent parton Fermi surface (FS) in the 2D limit~\cite{Xu2020}. On the contrary, another DMRG study~\cite{Sheng2021} and a self-consistent mean-field theory~\cite{Chen2021} prefer a stripe-ordered ground state.
Soon thereafter, using a state-of-the-art method, DMRG boosted by Gutzwiller projected wave functions~\cite{Jin2020_2}, two of the authors and their collaborators revisited this model~\cite{jin2022Bulletin}. The latest work revealed a nematic quantum spin-orbital liquid state, a critical stripy state that preserves SU(4) symmetry but breaks translational symmetry by doubling the unit cell along one of two primitive vectors. It was shown that the central charge of each stripe is $c = 3$, which is in quantitative agreement with the SU(4)$_1$ WZW CFT.
Moreover, it was found that the DMRG-obtained state can be well described by a ``single" Gutzwiller projected wave function with an emergent parton FS that consists of open orbits in the reciprocal space and indicates the nematicity. Such a ``unified" picture captures all the essential physics in both quasi-1D cylinders (with circumferences up to $L_y=8$) and the 2D limit, although the former exhibits a strong finite-size effect and even-odd discrepancy~\cite{jin2022Bulletin}.
Note that the newly proposed parton FS consists of open orbits in reciprocal space, which must undergo a Lifshitz transition from the originally proposed $\pi$-flux state, which has a closed parton FS~\cite{Xu2020}. 

In this work, we re-examine the SU(4) KK model on the triangular lattice by using the variational Monte Carlo (VMC) approach. The advantage of the VMC method is to reduce the impact of finite-size effects in 2D systems and to impose periodic boundary conditions (PBCs) in both directions. Thus, the previously proposed Gutzwiller projected wave function~\cite{jin2022Bulletin} can be carefully verified at large tori, up to a lattice size of $24\times{}24$. To this end, we have proposed five types of spin-orbital liquid states: a uniform $\pi$-flux state, three stripy states, and a plaquette state. After extensive numerical efforts, we have gathered strong evidence pointing to a critical stripy ground state, as revealed by the Gutzwiller-boosted DMRG~\cite{jin2022Bulletin}.  

The rest of this paper is organized as follows. We revisit the SU(4) KK model in Section~\ref{sec:sec2} to make it self-contained. Then five fermionic parton mean-field ansatzes have been proposed in Section~\ref{sec:sec3}. In Section~\ref{sec:sec4}, we search for the ground state with the help of VMC and the stochastic reconfiguration method. Section~\ref{sec:sec5} is devoted to summary.

\section{Model and symmetries}\label{sec:sec2}

The SU(4) KK model Hamiltonian on the triangular lattice is defined as
\begin{subequations}~\label{eq:H}
\begin{equation}
H=\frac{1}{2}\sum_{\langle i,j \rangle}\left(4\vec{S}_i\cdot \vec{S}_j+1\right)\left(4\vec{T}_i\cdot \vec{T}_j+1\right), 
\end{equation}
where $\langle i,j \rangle$ denotes an NN bond and $\vec{S}_i$ ($\vec{T}_i$) is the $S=1/2$ spin (orbital) vector at site $i$. These spin (orbital) vectors can be represented by introducing $\vec{S}_i=\frac{1}{2}\{\sigma^{x}_i,\sigma^{y}_i,\sigma^{z}_i\}$ ($\vec{T}_i=\frac{1}{2}\{\tau^{x}_i,\tau^{y}_i,\tau^{z}_i\}$), where $\sigma^{x,y,z}$ ($\tau^{x,y,z}$) are the standard Pauli matrices acting on the two-fold spin (orbital) indices. 
It is easy to verify that the Hamiltonian $H$ in Eq.~\eqref{eq:H} commutes with all the 15 generators of the SU(4) Lie group, $\lambda^{1},\lambda^{2},\cdots,\lambda^{15}$, that are linear combinations of $\{\vec{\sigma},\vec{\tau},\vec{\sigma}\otimes{}\vec{\tau} \}$. Therefore the SU(4) symmetry is respected by the model Hamiltonian.

The SU(4) symmetry becomes more transparent from the point of view that the $2\times{}2=4$ spin-orbital degrees of freedom at each site can be treated as a pseudospin. To demonstrate this, we introduce a 15-dimensional vector $\vec{\lambda}_{i}\equiv\{\lambda_{i}^{1},\lambda_{i}^{2},\cdots,\lambda_{i}^{15}\}$ whose components are 15 SU(4) generators. Then the model Hamiltonian can be rewritten in terms of $\vec{\lambda}_{i}$ as
\begin{equation}
H=\sum_{\langle i,j \rangle}\left(\vec{\lambda}_i\cdot \vec{\lambda}_j+\frac{1}{2}\right).
\end{equation}
\end{subequations}
Note that these generators, $\lambda_{i}^{\mu}$'s, have been normalized with $\mbox{tr}(\lambda_{i}^{\mu}\lambda_{j}^{\nu})=2\delta_{ij}\delta_{\mu\nu}$~\cite{Georgi}. So the Hamiltonian in Eq.~\eqref{eq:H} can be interpreted as an SU(4) AFM Heisenberg model, which is invariant under the global SU(4) pseudospin-rotation symmetry.

In addition to the SU(4) symmetry, this model is also symmetric with respect to the spatial symmetries of the triangular lattice, including the lattice translation symmetries $T_{1,2}$, the mirror symmetry $M$, and the sixfold rotation symmetry $C_6$. Furthermore, as a spin-orbital system, this model naturally preserves the time-reversal symmetry.

\section{Parton construction and mean-field ansatzes}\label{sec:sec3}

In this section, we shall first use the Gutzwiller projection to construct a trial wave function for solving the Hamiltonian $H$ in Eq.~\eqref{eq:H}. Given a specific variational wave function $|\Psi_{trial}\rangle$, one can calculate the ground-state energy $E$,
\begin{equation*}
E=\frac{\langle\Psi_{trial}{}|H|\Psi_{trial}\rangle}{\langle\Psi_{trial}|\Psi_{trial}\rangle},
\end{equation*}
by using the standard Monte Carlo method. This trial wave function can then be optimized by minimizing the energy $E$.

To construct such a trial Gutzwiller projected state, we introduce four flavors of fermionic partons at each site $i$, $f^{\dagger}_{i\alpha}$ for $\alpha=1,2,3,4$ ($f_{i\alpha}$ the annihilation operators), where $\alpha$ denotes the four pseudospin (spin-orbital) degrees of freedom mentioned above.
The SU(4) pseudospin operators can now be expressed in terms of these fermionic partons as 
\begin{equation}\label{eq:parton}
\lambda^{\mu}_i= f_i^\dag \lambda^{\mu} f_i, \quad{\rm~for~}\mu=1,\cdots,15,
\end{equation}
where $f_{i}\equiv(f_{i1},f_{i2},f_{i3},f_{i4})^{T}$ is a four-component vector. Instead of $4$ spin-orbital states, these four flavors of fermions will generate $2^4=16$ states in Fock space at each local site. To restore the physical Hilbert space, a single occupancy constraint of $\sum_{\alpha=1}^4 f_{i\alpha}^\dag f_{\alpha}=1$ was imposed on each site, e.g. the Gutzwiller projection was implemented.

In this parton representation in Eq.~\eqref{eq:parton}, the inner product of two pseudospins $\vec{\lambda}_i\cdot \vec{\lambda}_j$ can be written in terms of four-fermion interactions:
\begin{equation}\label{eq:4fermion}
\begin{split}
\vec{\lambda}_i\cdot \vec{\lambda}_j = & - 2\sum_{\alpha,\beta=1}^{4}\left({}f_{i\alpha}^{\dag}f_{j\alpha}f_{j\beta}^{\dag}f_{i\beta} + \frac{1}{4}{}f_{i\alpha}^{\dag}f_{i\alpha}f_{j\beta}^{\dag}f_{j\beta}\right)\\
& + \sum_{\alpha=1}^{4}\left(f^\dag_{i\alpha}f_{i\alpha}+f^\dag_{j\alpha}f_{j\alpha}\right) 
\end{split}.  
\end{equation}
Here the identity $\vec{\lambda}_{ab}\cdot\vec{\lambda}_{cd}=2\delta_{ad}\delta_{bc}-\frac{1}{2}\delta_{ab}\delta_{cd}$ has been used. Thus, the model Hamiltonian $H$ in Eq.~\eqref{eq:H} can be recast as follows:
\begin{equation}\label{eq:H-parton}
H = \sum_{\langle{}ij\rangle} \left(-\sum_{\alpha,\beta=1}^{4}{}f_{i\alpha}^{\dag}f_{j\alpha}f_{j\beta}^{\dag}f_{i\beta} + 1\right).
\end{equation}
Note that the single-occupancy condition was imposed in the derivation of the last equation.
The four-fermion term can be decoupled by introducing the mean-field parameters on NN bonds $\langle{}ij\rangle$ as 
\begin{equation}
\chi_{ij}=\chi_{ji}^{\ast}=\sum_{\alpha=1}^4 \langle f_{i\alpha}^\dag f_{j\alpha} \rangle.
\end{equation}
So we get a quadratic mean-field Hamiltonian 
\begin{equation}\label{eq:HMF0}
H_{MF}=\sum_{\langle{}ij\rangle}\sum_{\alpha=1}^{4}\chi_{ij}f_{j\alpha}^{\dagger}f_{i\alpha}+h.c.,
\end{equation}
which preserves the SU(4) symmetry and can be easily diagonalized now.

In fact, $H_{MF}$ in Eq.~\eqref{eq:HMF0} can be treated as a ``variational Hamiltonian'' or an ``effective Hamiltonian'' on which $\{\chi_{ij}\}$ is a set of variational parameters to be determined. Taking into account the single occupancy constraint at the mean, i.e. $\sum_{\alpha=1}^4 \langle{}f_{i\alpha}^\dag f_{i\alpha}\rangle=1$, one can find a ground state $|\Psi_{MF}\rangle$ for the mean-field Hamiltonian $H_{MF}$ at $1/4$ filling. Then the trial wave function $|\Psi_{trial}\rangle$ can be constructed by performing the Gutzwiller projection on $|\Psi_{MF}\rangle$:
\begin{equation}
|\Psi_{trial}\rangle = P_{G}|\Psi_{MF}\rangle.\label{eq:trial}
\end{equation}
Here $P_{G}$ is the Gutzwiller projector, which removes all non-single-occupied components to locally enforce the single-occupancy constraint.

The set of parameters, $\{\chi_{ij}\}$, together with the form of $H_{MF}$ in Eq.~\eqref{eq:HMF0}, is called ``{\it{}mean-field Ansatz}".
As long as a mean-field Ansatz is given, one can evaluate the energy $E(\{\chi_{ij}\})$ by using the standard VMC method, and minimize $E(\{\chi_{ij}\})$ to obtain an optimal set of parameters $\{\chi_{ij}\}$. Then, all the correlation functions can be computed with such an optimized wave function. It is worth mentioning that the mean-field parameters $\chi_{ij}$ can be determined up to an overall positive factor via the ground state energy optimization.
In the remaining part of this section, we shall discuss several typical examples of SU(4)-symmetric mean-field Ansatzes which will be utilized to construct trial wave functions.  

\begin{figure}[tb]
\centering
\subfigure[]{
\includegraphics[width=0.95\linewidth]{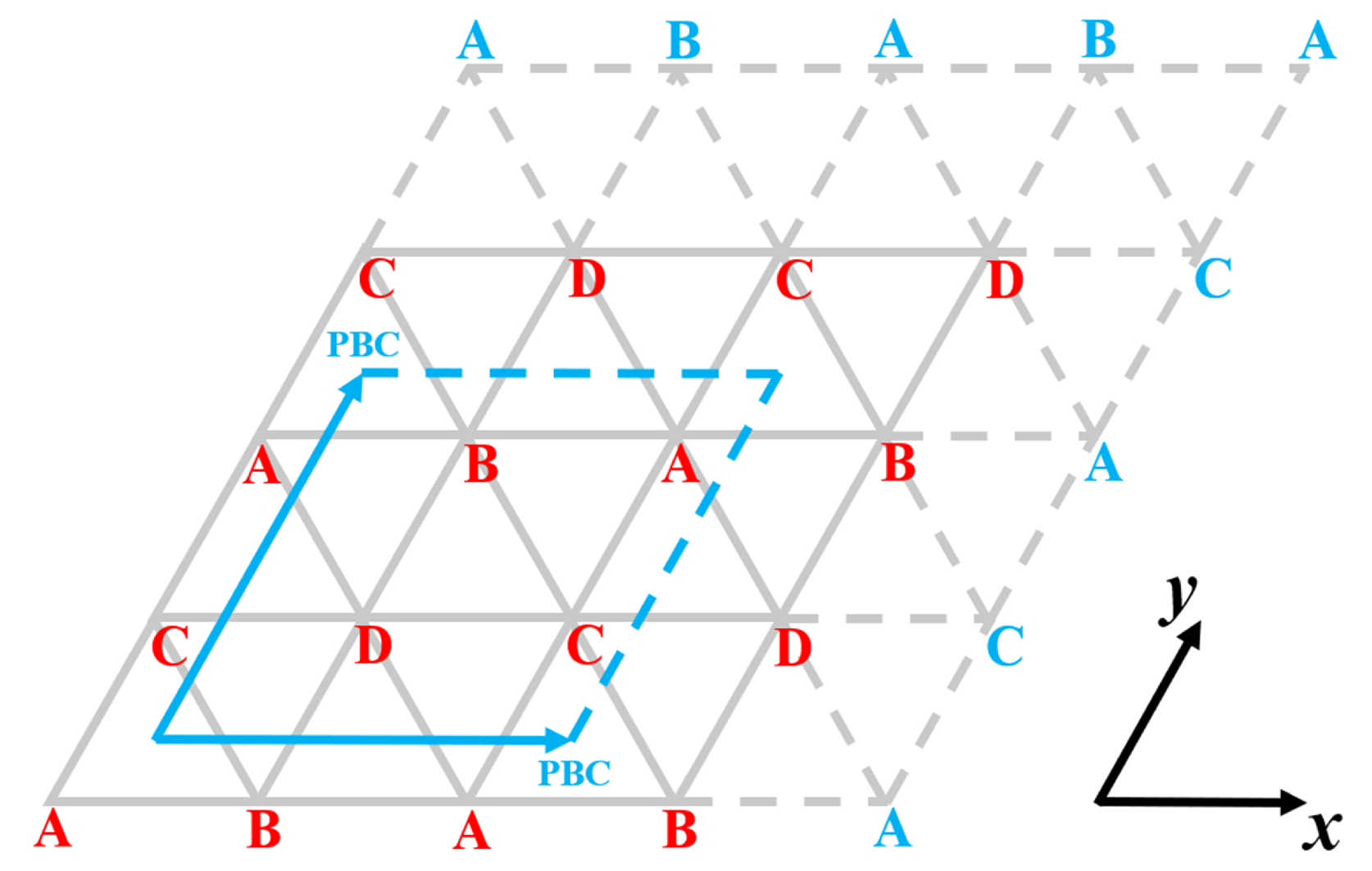}
}
\subfigure[]{
\includegraphics[width=0.6\linewidth]{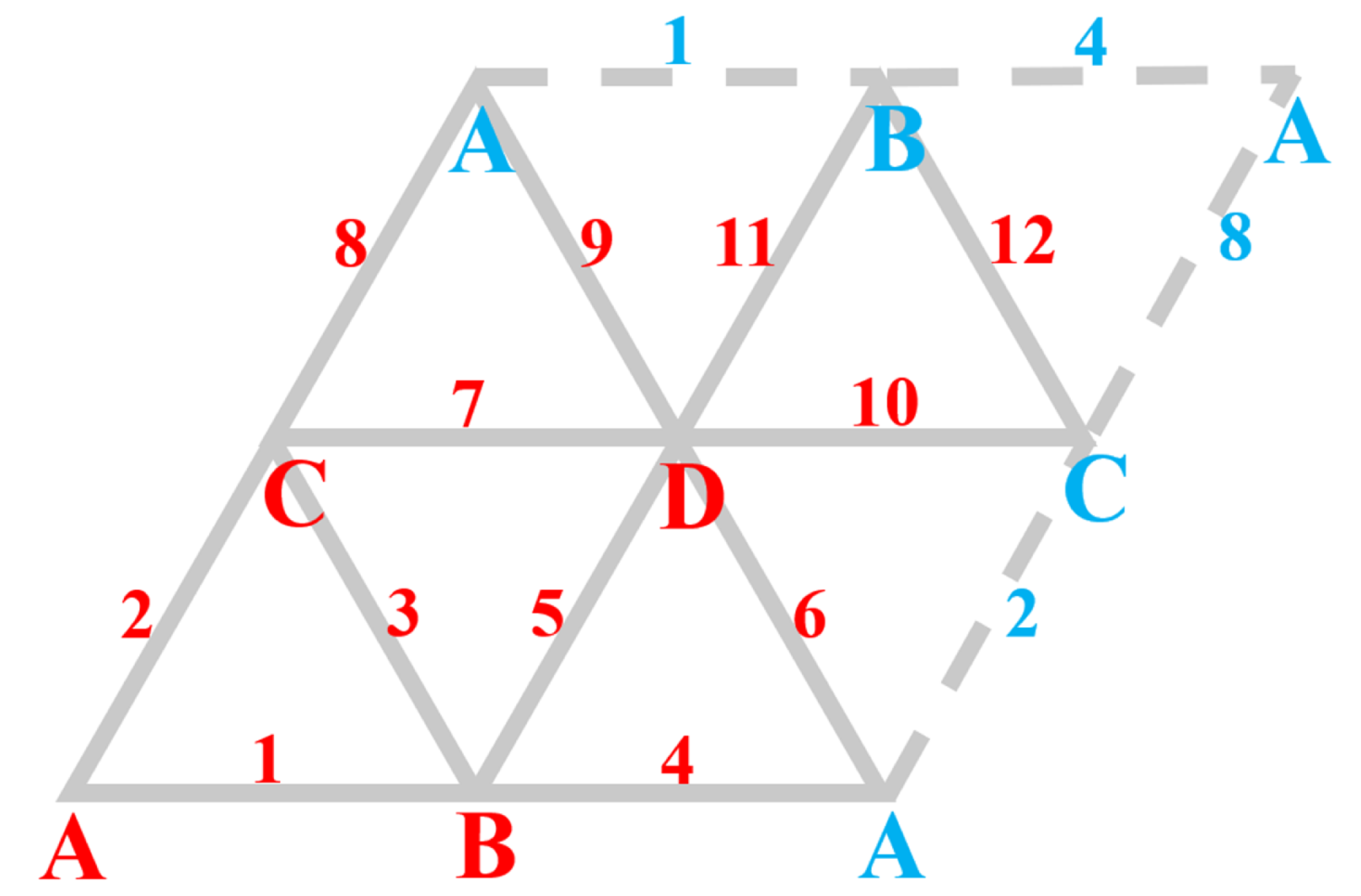}
}
\caption{For a $2\times{}2$ VBC on a triangular lattice, the unit cell has been enlarged to include four sublattices denoted by A, B, C, and D, respectively. (a) The XC torus geometry and the periodic boundary condition (PBC) are shown by dashed lines. (b) The 12 types of NN bonds in the enlarged unit cell.} ~\label{fig:fig1}
\end{figure}

\begin{figure*}[tb]
\centering
\includegraphics[width=\linewidth]{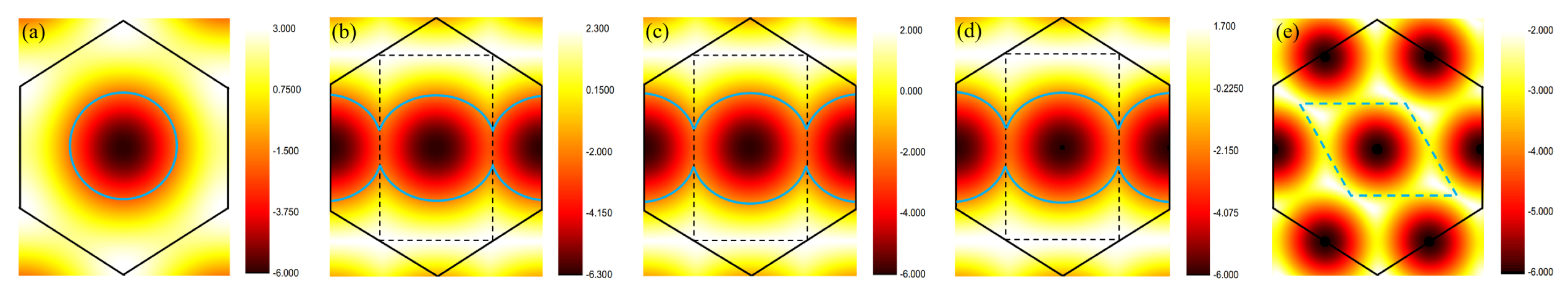}
\caption{The parton energy contour plot in the first (unfolded) Brillouin zone (BZ) for (a) the uniform $\pi$-flux state [Eq.~\eqref{eq:piflux}], (b) the stripe-I state [$\delta_1=0.15$ in Eq.~\eqref{eq:stripyI}], (c) the stripe-II state [$\delta_1=0.15$ and $\epsilon=0$ in Eq.~\eqref{eq:stripyII}], (d) the stripe-III state [$\delta_1=0.15$ in Eq.~\eqref{eq:stripyIII}], and (e) the plaquette state [$\delta_1=\delta_2=0.22$ in Eq.~\eqref{eq:plaq}], where the parameter $t=1$ be set. The dashed lines in (b)-(e) enclose the folded BZ. At $1/4$ filling, (a) the uniform $\pi$-flux state has a closed Fermi surface (FS), (b)-(d) stripe states allow an open FS in the BZ, while (e) the plaquette state is a parton band insulator. } ~\label{fig:fermisurface}
\end{figure*}

{\bf{}Uniform $\pi$-flux state}: Two of the simplest mean-field ansatzes are uniform zero-flux or $\pi$-flux states. Since the latter always has a lower energy than the former, we will skip the higher-energy zero-flux state and discuss only the uniform $\pi$-flux state.
The uniform $\pi$-flux state is governed by the mean-field Hamiltonian as follows,
\begin{equation}
H_{\Delta=\pi}=-t\sum_{\alpha=1}^4\sum_{\langle ij \rangle}(f_{i\alpha}^\dag f_{j\alpha} + h.c.), ~\label{eq:piflux}    
\end{equation}
which is obtained by setting $\chi_{ij}=-t\,\,(t>0)$ on all the NN bonds in Eq.~\eqref{eq:HMF0}. 
In each elementary triangle, $\chi_{ij}$ will pick up a $\pi$-flux that gives a name to the $\pi$-flux state proposed in Ref.~\cite{Xu2020}. The parton FS at $1/4$ filling is plotted in Fig.~\ref{fig:fermisurface}(a), which is almost a circle.

As an instability of the uniform $\pi$-flux state, a ``stripy state" has recently been proposed~\cite{jin2022Bulletin}. The stripy state breaks the lattice translational symmetry by doubling the unit cell along a certain lattice vector direction. In this work, we consider a more generic class of valence-bond-crystal (VBC) states. These VBC states have an enlarged $2\times{}2$ unit cell, as illustrated in Fig.~\ref{fig:fig1}(a). One can obtain stripy states or the uniform $\pi$-flux state by restoring the lattice translational symmetry along one or two lattice vector directions.

{\bf{}Generic $2\times{}2$ VBC states:} Since the coordination number is $z=6$ on a triangular lattice. Each unit cell in a $2\times{}2$ VBC state has $2\times{}2\times{}6/2=12$ NN bonds, as labeled by $n=1,2,\cdots,12$ in Fig.~\ref{fig:fig1}(b). 
Thus, a generic VBC state can be depicted by 12 mean-field parameters $\chi_{ij}$ within one unit cell as 
\begin{equation}
\chi_{ij} = -t_{n}, \label{eq:chiVBC}   
\end{equation}
such that the corresponding mean-field Hamiltonian takes the form of
\begin{equation}
H_{VBC}=-\sum_{\alpha=1}^4\sum_{n=1}^{12}\sum_{\langle ij \rangle_n}t_n\left(f_{i\alpha}^\dag f_{j\alpha} + h.c.\right), ~\label{eq:HVBC}
\end{equation}
where $\langle{}ij\rangle_{n}$ denotes an type-$n$ NN bond as shown in Fig.~\ref{fig:fig1}(b). As mentioned above, stripy states can be obtained by restoring the lattice translational symmetry along either the $x$- or the $y$-direction.

\begin{table*}[tb]
\centering
\caption{Five types of mean-field ansatzes. Here $t_n>0$ on all the NN bonds $\langle{ij}\rangle_{n=1,\cdots,12}$, suggesting that $\pi$-flux states are energetically favorable. $T_x(T_y)$ shifts the system along the $x(y)$- direction by a lattice constant. }\label{tab:t1}
\renewcommand\arraystretch{2.0}
\setlength{\tabcolsep}{1.5ex}
\begin{tabular}{c|c|c|c|c}
\hline\hline
\begin{minipage}[b]{0.4\columnwidth}\raisebox{-.5\height}{\includegraphics[width=\linewidth]{p1b.pdf}}\end{minipage}
&\text{schematic bond strength}
&\text{mean-field ansatz}
&\text{unit cell} & \text{symmetry}\\ \hline
\text{\makecell{(a) uniform $\pi$-flux  state}}
& \begin{minipage}[b]{0.7\columnwidth}\raisebox{-.5\height}{\includegraphics[width=\linewidth]{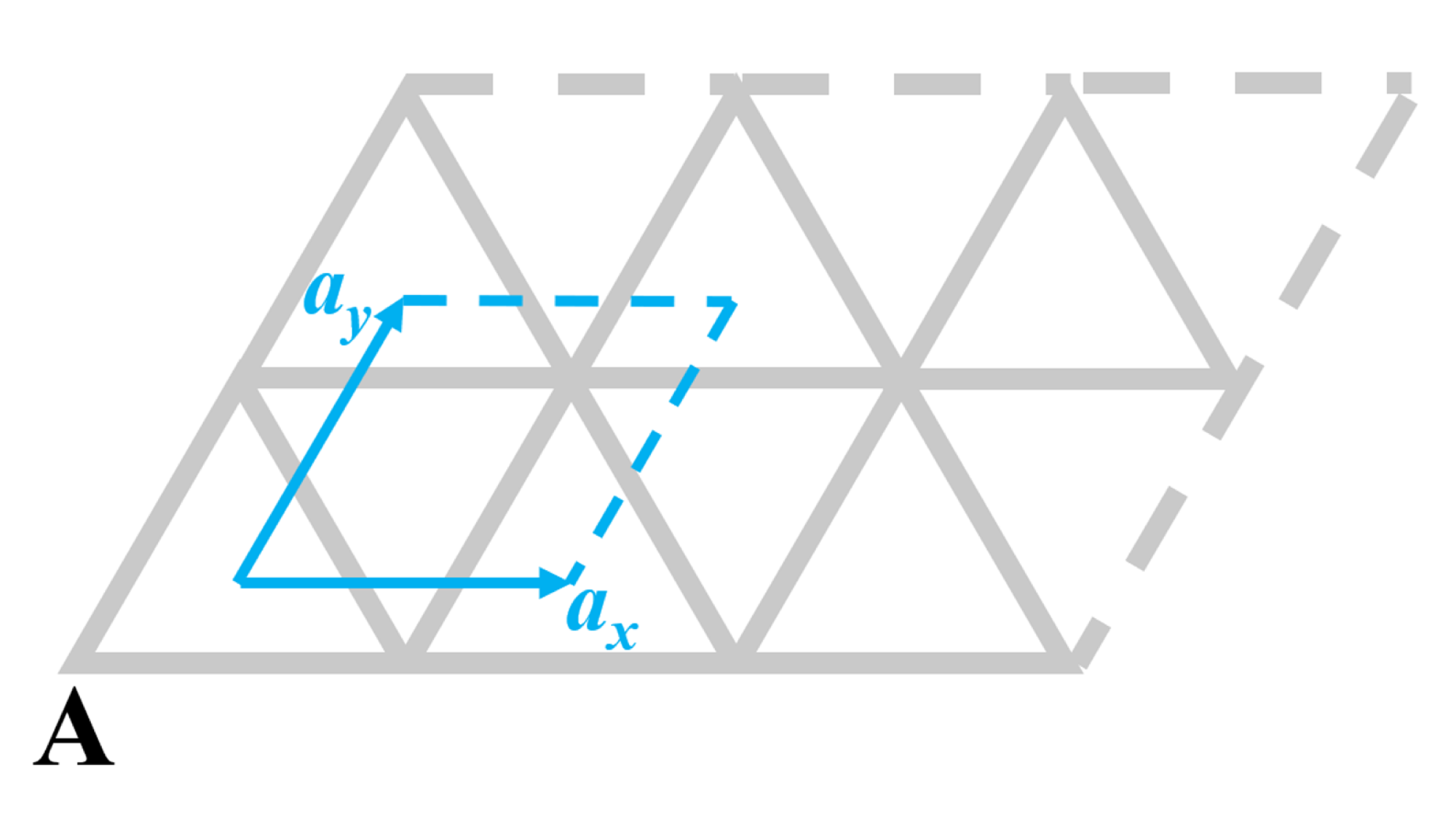}}\end{minipage}
& \makecell{$\chi_{ij}=-t$\\ \text{see Eq.~\eqref{eq:piflux}} }
& $1\times{}1$ & \makecell{$T_{x},T_{y}$\\ $D_{6}$} \\ \hline

\text{(b) stripe-I state}
& \begin{minipage}[b]{0.7\columnwidth}\raisebox{-.5\height}{\includegraphics[width=\linewidth]{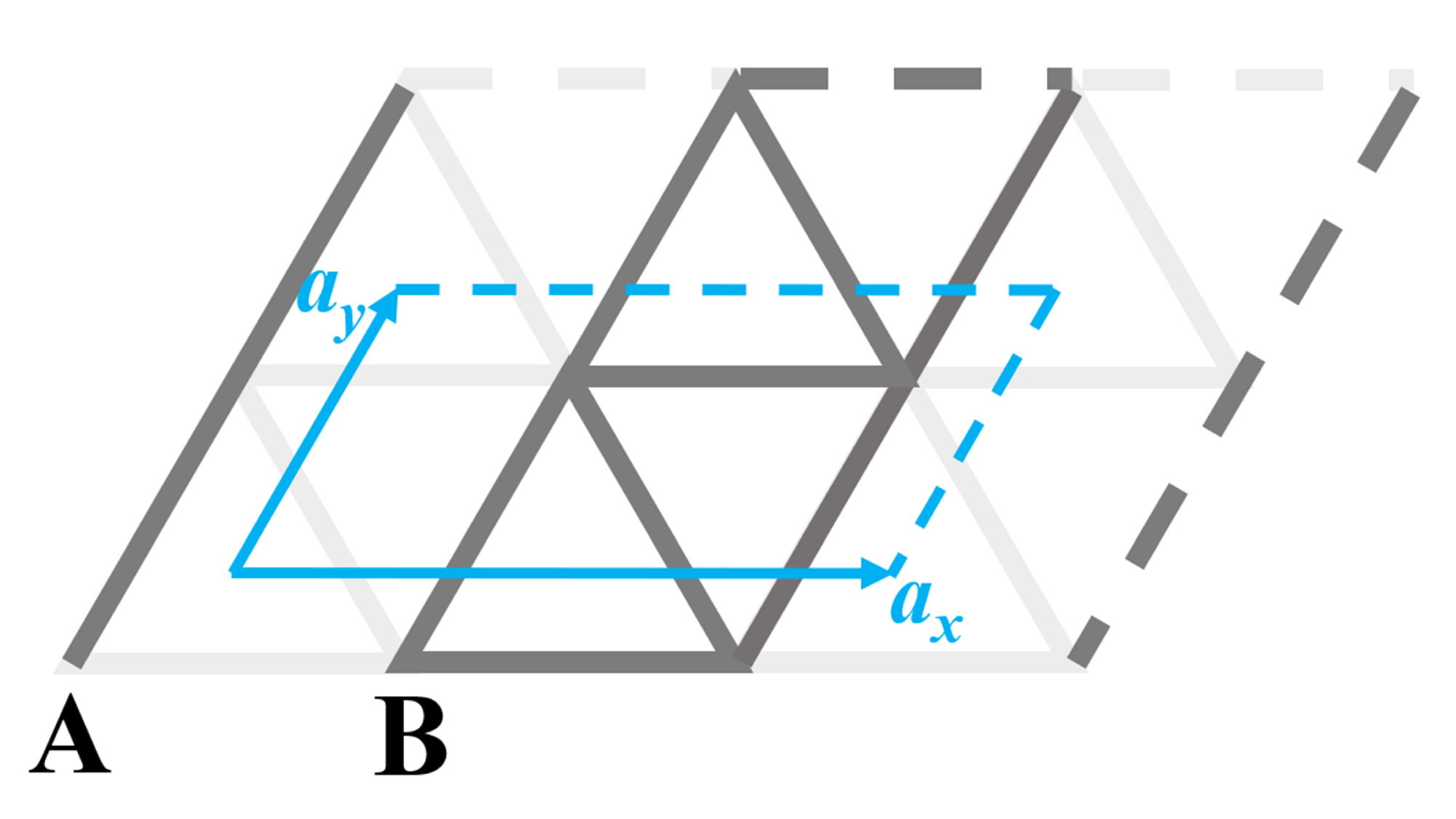}}\end{minipage}
& \text{\makecell{$t_{n+6}=t_{n}|_{(n=1,\cdots,6)}$\\$t_2=t_4=t_5$\\ see Eq.~\eqref{eq:stripyI}} } 
& $2\times{}1$ & \makecell{$T_{x}^2,T_y$\\$C_{2}$} \\ \hline

\text{(c) stripe-II state}
& \begin{minipage}[b]{0.7\columnwidth}\raisebox{-.5\height}{\includegraphics[width=\linewidth]{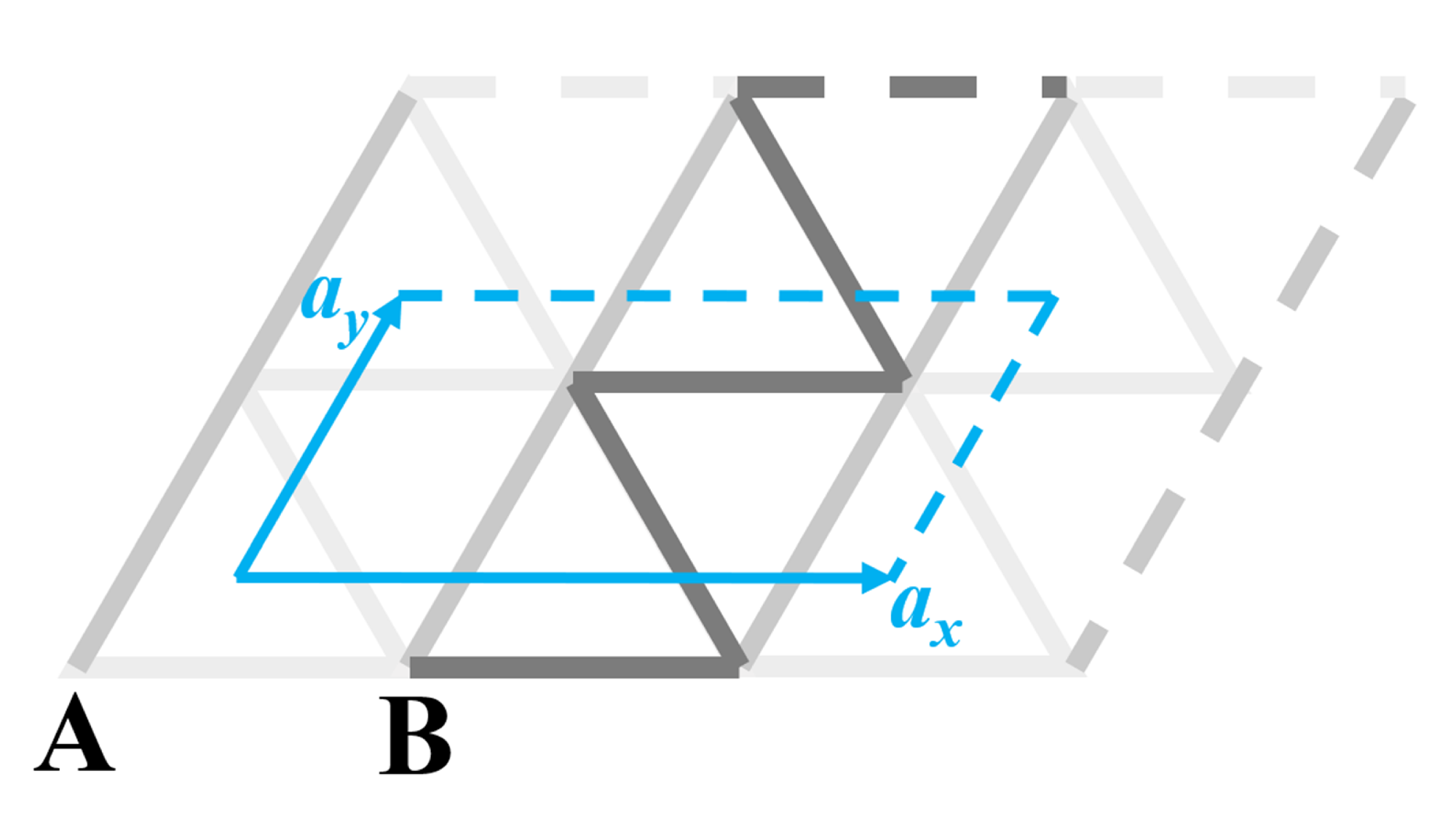}}\end{minipage}
& \text{\makecell{$t_{n+6}=t_{n}|_{(n=1,\cdots,6)}$\\$t_2=t_5$ \\ see Eq.~\eqref{eq:stripyII} }}
& $2\times{}1$ & \makecell{$T_{x}^2,T_y$\\$C_{2}$} \\ \hline

\text{(d) stripe-III state}
& \begin{minipage}[b]{0.7\columnwidth}\raisebox{-.5\height}{\includegraphics[width=\linewidth]{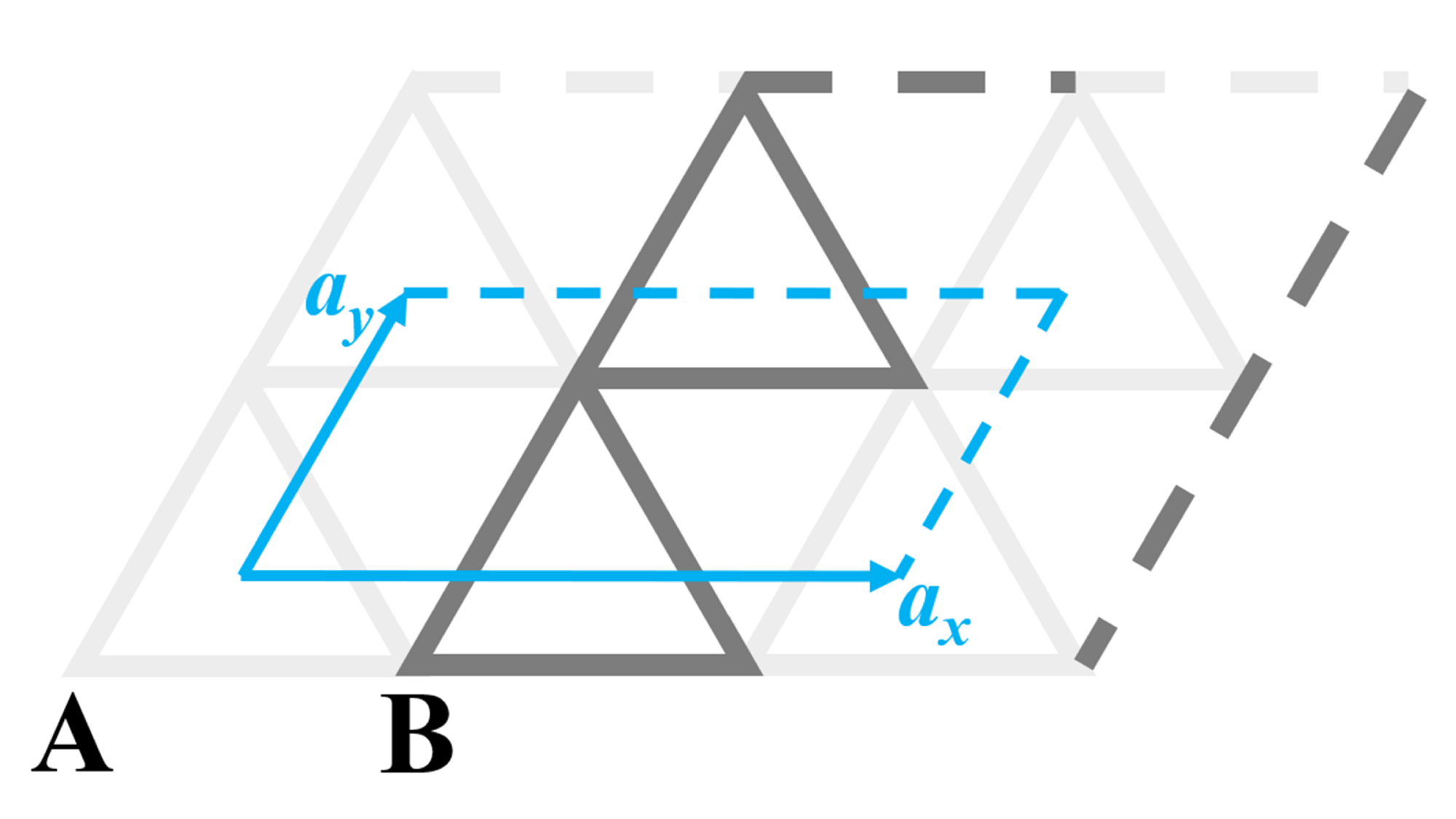}}\end{minipage}
& \text{\makecell{$t_{n+6}=t_{n}|_{(n=1,\cdots,6)}$\\$t_1=t_2$ \\ $t_4=t_5$\\ see Eq.~\eqref{eq:stripyIII} }}
& $2\times{}1$ & \makecell{$T_{x}^2,T_y$\\$C_{2}$} \\ \hline

\text{(e) plaquette state}
& \begin{minipage}[b]{0.8\columnwidth}\raisebox{-.5\height}{\includegraphics[width=\linewidth]{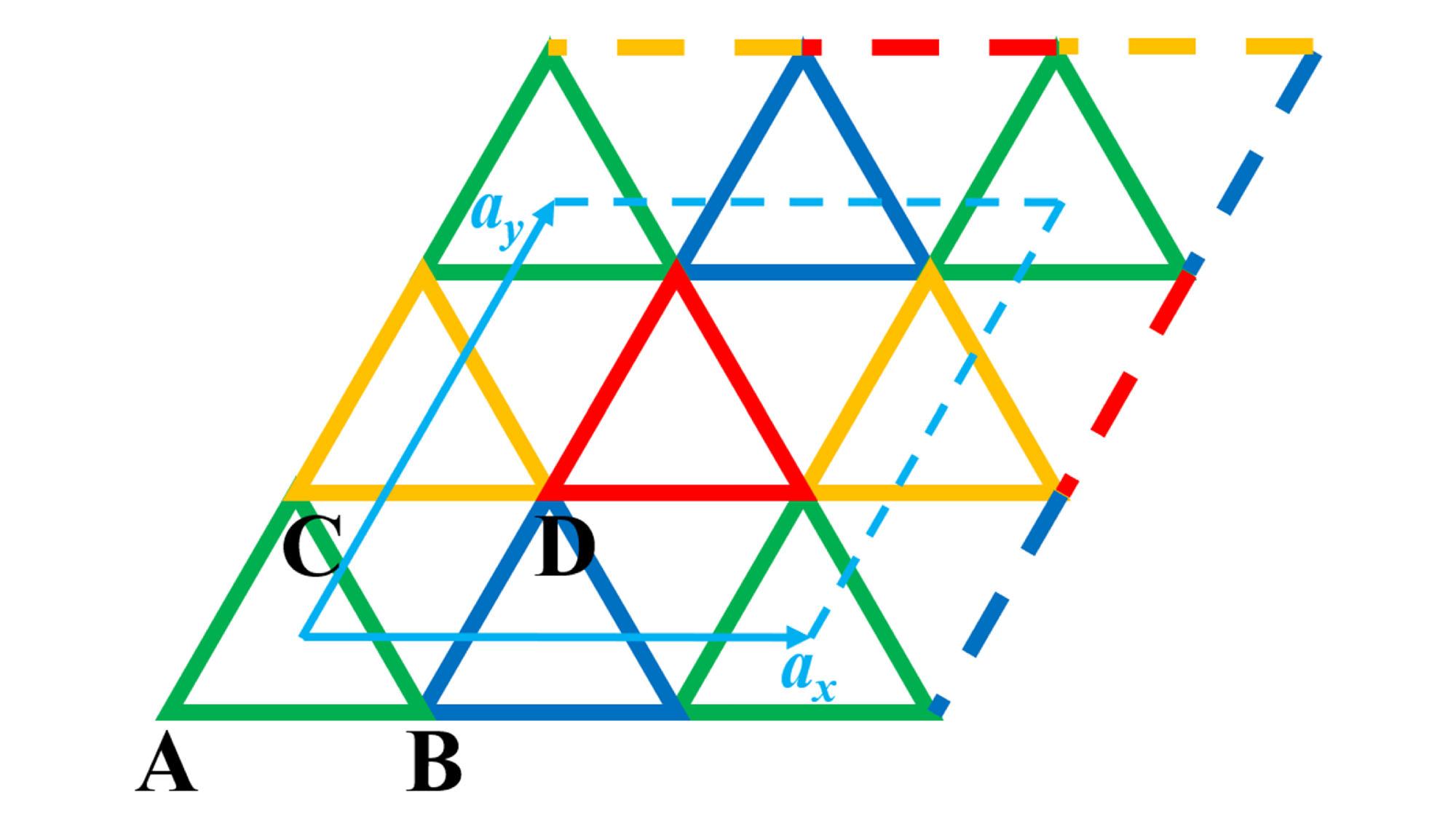}}\end{minipage}
& \text{\makecell{$t_1=t_2=t_3$\\$t_4=t_5=t_6$ \\ $t_7=t_8=t_9$\\$t_{10}=t_{11}=t_{12}$\\$t_{1}+t_{10}=t_{4}+t_{7}$\\ see Eq.~\eqref{eq:plaq} }}
& $2\times{}2$ & $T_{x}^2,T_{y}^2$ \\ 
\hline\hline
\end{tabular}
\end{table*}

{\bf{}$C_2$- stripe states:} From a symmetry point of view, a stripy state will break the lattice translational symmetry in one direction while preserving the other. The $D_{6}$ rotational symmetry of the lattice will be broken simultaneously. The resulting stripy state may or may not remain a reduced $C_2$ rotational symmetry of the original $D_6$ symmetry. In particular, we are interested in the stripy states exhibiting the $C_2$ rotational symmetry, dubbed the ``$C_2$- stripe states".

Without loss of generality, we focus on stripy states that only break the translational symmetry along the $x$-direction (associated with symmetry generator $T_{x}$). 
The unit cell is of $2\times{}1$ and consists of 6 NN bonds. Therefore, there are only two sublattices, e.g., $\mbox{A}=\mbox{C}$ and $\mbox{B}=\mbox{D}$ in Fig.~\ref{fig:fig1}(b). The 12 values of $\chi_{ij}$, say, $t_{n}$ ($n=1,2,\cdots,12$) defined according to the labeling in Fig.~\ref{fig:fig1}(b) are reduced to 6 values in such a $2\times{}1$ VBC state. Namely, the constraint $t_{n}=t_{n+6}$ ($n=1,2,3,4,5,6$) should be imposed on Eqs.~\eqref{eq:chiVBC} and \eqref{eq:HVBC}.
Introduce a lattice site labeling scheme,
\begin{equation*}
i=(x_{i},y_{i})=x_{i}\hat{x}+y_{i}\hat{y},
\end{equation*}
we find that for a $2\times{}1$ stripy state, $\chi_{ij}$ must satisfy on NN bonds:
\begin{equation*}
\chi_{i,i+\hat{x}}=-t_{1(4)},\,\,\, \chi_{i,i+\hat{y}}=-t_{2(5)},\,\,\, \chi_{i,i+\hat{x}-\hat{y}}=-t_{3(6)},\,\,\, i\in\mbox{A (B)}.
\end{equation*}
Now we go further to consider $C_2$- stripe states that respect the reduced $C_2$ rotational symmetry around the $\hat{x}-\frac{1}{2}\hat{y}$ axis, giving rise to $t_1=t_3$ and $t_4=t_6$. Therefore, the mean-field ansatz will take a simplified form on the 6 NN bonds in the $2\times{}1$ unit cell, as 
\begin{equation}\label{eq:stripeC2}
\chi_{i,i+\hat{x}}=\chi_{i,i+\hat{x}-\hat{y}}=-t_{1(4)},\,\,\, \chi_{i,i+\hat{y}}=-t_{2(5)},\,\,\, i\in\mbox{A (B)}. 
\end{equation}
Note that due to symmetry there are only 4 different values of $\chi_{ij}$, i.e. $t_1$, $t_2$, $t_4$ and $t_5$. Moreover, such a $C_2$-stripe state is characterized by at most three independent parameters up to a positive total factor.
In the following, we will discuss three types of $C_2$- stripe states, called ``stripe-I, -II, and -III states'', which have been found by the VMC optimization to be local minima in the full parameter space for $2\times{}2$ VBC states, $\{ t_{1},t_{2},\cdots,t_{12} \}$. 

{\bf{}Stripe-I state:} This type of states can be achieved by setting
\begin{subequations}\label{eq:stripyI}
\begin{equation}
t_2=t_4=t_5 
\end{equation}
in Eq.~\eqref{eq:stripeC2}. The mean-field ansatz for such a stripy state can be simplified as follows:
\begin{equation}
\chi_{ij} = -t(1+r_{ij}\delta_{1}),
\end{equation}
where $r_{ij}=\pm{}1$ is an integer defined by
\begin{equation}\label{eq:rij}
r_{ij}=\delta_{x_{i},x_{j}}+(-1)^{x_{i}}(1-\delta_{x_{i},x_{j}}),
\end{equation}
\end{subequations}
$t=(t_1+t_4)/2$ serves as an overall factor, $\delta_{1}=(t_4-t_1)/(t_4+t_1)$ is the contrast of the stripe, and $x_{i}\le{}x_{j}$ is assumed for NN bond $\langle{}ij\rangle$.
Note that the stripe-I state is exactly equivalent to the trial wave function used to initialize the DMRG calculations in Ref.~\cite{jin2022Bulletin}.

{\bf{}Stripe-II state:} Imposing the constraint
\begin{subequations}\label{eq:stripyII}
\begin{equation}
t_2=t_5
\end{equation}
in Eq.~\eqref{eq:stripeC2} gives rise to the stripe-II state. The mean-field ansatz for such a state reads
\begin{equation}
\chi_{ij} = -t\left[1+(-1)^{x_{i}}\delta_{1}\right](1-\delta_{x_{i},x_{j}})-t(1+\epsilon)\delta_{x_{i},x_{j}},
\end{equation}
\end{subequations}
where $t=(t_1+t_4)/2$ is an overall factor, $\delta_{1}=(t_4-t_1)/(t_4+t_1)$ reflects the stripe contrast,  $\epsilon=t_{2}/t-1$, and $x_{i}\le{}x_{j}$ has been assumed for NN bond $\langle{}ij\rangle$. When $\epsilon=\delta_1$, the stripe-II state becomes a stripe-I state.

{\bf{}Stipe-III state:} Such a state can be achieved by setting
\begin{subequations}\label{eq:stripyIII}
\begin{equation}
t_1=t_2\,\mbox{ and }\,t_4=t_5
\end{equation}
in Eq.~\eqref{eq:stripeC2}. The corresponding mean-field ansatz reads
\begin{equation}
\chi_{ij}  = -t\left[1+(-1)^{x_{i}}\delta_{1}\right],
\end{equation}
\end{subequations}
where $t=(t_1+t_4)/2$ and $\delta_{1}=(t_{4}-t_{1})/(t_4+t_1)$, and $x_{i}\le{}x_{j}$ has been assumed for NN bond $\langle{}ij\rangle$.

{\bf{}Plaquette state:} In addition to the uniform $\pi$-flux state and the three $C_2$- stripe states, we also consider the ``plaquette state", which is given by ``perfect triangle conditions":
\begin{subequations}\label{eq:plaq}
\begin{equation}
\begin{split}
t_{1}&=t_{2}=t_{3},\\
t_{4}&=t_{5}=t_{6},\\
t_{7}&=t_{8}=t_{9},\\
t_{10}&=t_{11}=t_{12},
\end{split}
\end{equation}
and the constraint
\begin{equation}
t_{1}+t_{10}=t_{4}+t_{7}.
\end{equation}
Thus, the mean-field ansatz for such a plaquette state takes a three-parameter form of
\begin{equation}
\chi_{ij}=-t[1+(-1)^{x_i}\delta_1+(-1)^{\min\{y_{i},y_{j}\}}\delta_2].
\end{equation}
where $x_{i}\le{}x_{j}$ has been assumed for NN bond $\langle{}ij\rangle$, and the three parameters $t$, $\delta_{1}$, as well as $\delta_{2}$ are determined by
\begin{equation}
\begin{split}
t_{1}+t_{10}&=t_{4}+t_{7}=2t,\\
t_{1}&=t(1+\delta_1+\delta_2),\\
t_{4}&=t(1+\delta_1-\delta_2).
\end{split}
\end{equation}
\end{subequations}
It is worth noting that the parameter $t$ for these five states that are defined in Eqs.~\eqref{eq:piflux}, \eqref{eq:stripyI}, \eqref{eq:stripyII}, \eqref{eq:stripyIII} and \eqref{eq:plaq} respectively can be unified as the average of the 12 $t_n$'s in Eq.~\eqref{eq:HVBC},
\begin{equation}\label{eq:t}
t=\frac{1}{12}\sum_{n=1}^{12}t_{n}.    
\end{equation}

Before the end of this section, we summarize five types of typical parton states in Table~\ref{tab:t1}, including (a) the uniform $\pi$-flux state, (b) the stripe-I state, (c) the stripe-II state, (d) the stripe-III state, and (e) the plaquette state. With $1/4$ parton filling, all stripy states allow an open parton FS when the stripe contrast exceeds a critical value, $\delta_1>\delta_{c}$ (with $\delta_c\sim{}0.1$), as shown in Fig.~\ref{fig:fermisurface}. However, the plaquette state is always a band insulator of fermionic partons.

\section{VMC calculations and results}\label{sec:sec4}

We have performed VMC calculations on various $L_x \times L_y$ XC lattices up to $L_x=L_y=24$ [see Fig.~\ref{fig:fig1}(a)], on which PBCs for SU(4) pseudospins have been imposed along both $x$- and $y$-directions. Thus, the parton boundary conditions can be either PBCs or antiperiodic boundary conditions (APBCs). Indeed, we have found that the system always has a lower variational energy when the parton boundary conditions have been chosen to be APBCs rather than PBCs along both $x$- and $y$- directions.
 
First, we have searched for local energy minima with generic $2\times{}2$ VBC states given in Eq.~\eqref{eq:HVBC}, parameterized by the set of $\{t_{1},t_{2},\cdots,t_{12}\}$. 
To handle such a large parameter space, traditional methods such as steepest descent or Hessian matrix construction are computationally expensive and inefficient for optimizing the mean-field parameters. Therefore, a stochastic reconfiguration (SR) method has been exploited to optimize the parameters~\cite{Sorella1998,Sorella2000}. 
In our calculations, a dynamic step length for the variational parameters $t_n$, $\zeta$, has been attempted during the VMC simulations, instead of a constant step length.  At the beginning of optimization, a relatively large step length modulation, $\zeta \sim 5\times 10^{-2}$, is chosen to avoid trapping $E$ in local minima. Then we gradually decrease the modulation of $\zeta$ and in the last $10^{2}$ Monte Carlo steps, the step length is refined as $\zeta \sim 10^{-4}$. More details on the SR method can be found in Appendix A.

By performing extensive VMC computations initialized with a large number of sets of random $\{t_1,\cdots,t_{12}\}$, we can draw two conclusions as follows:

\newtheorem{theorem}{Conclusion}
\begin{theorem} 
All the energetically favorable states, found at local energy minima, correspond to a $\pi$-flux configuration, i.e., $t_{n}>0$ for all the $n=1,\cdots,12$ in Eq.~\eqref{eq:chiVBC}.
\end{theorem}

\begin{theorem} 
Within the numerical error bars, each local energy minimum found by the 12-parameter VMC optimizations can be identified as either the uniform $\pi$-flux state, one of the three $C_2$ stripy states, or a plaquette state (see Table~\ref{tab:t1}).
\end{theorem}

Based on these two results, we have performed further VMC calculations on $L_{x}=L_{y}=L$ tori using the five promising parton mean-field ansatzes listed in Table ~\ref{tab:t1}. Explicitly, in addition to the 12-parameter VMC approach, we have also applied the parameter constraints defined in Eqs. ~\eqref{eq:stripyI}, \eqref{eq:stripyII}, \eqref{eq:stripyIII}, and \eqref{eq:plaq} to our VMC calculations. Indeed, we find that all these simulations converge to energy minima that share the same energy with those obtained by the 12-parameter VMC optimization, supporting Conclusion 2. More details can be found in Appendix B.

\begin{figure*}[htb]
\centering
\subfigure[]{
\includegraphics[width=0.48\linewidth]{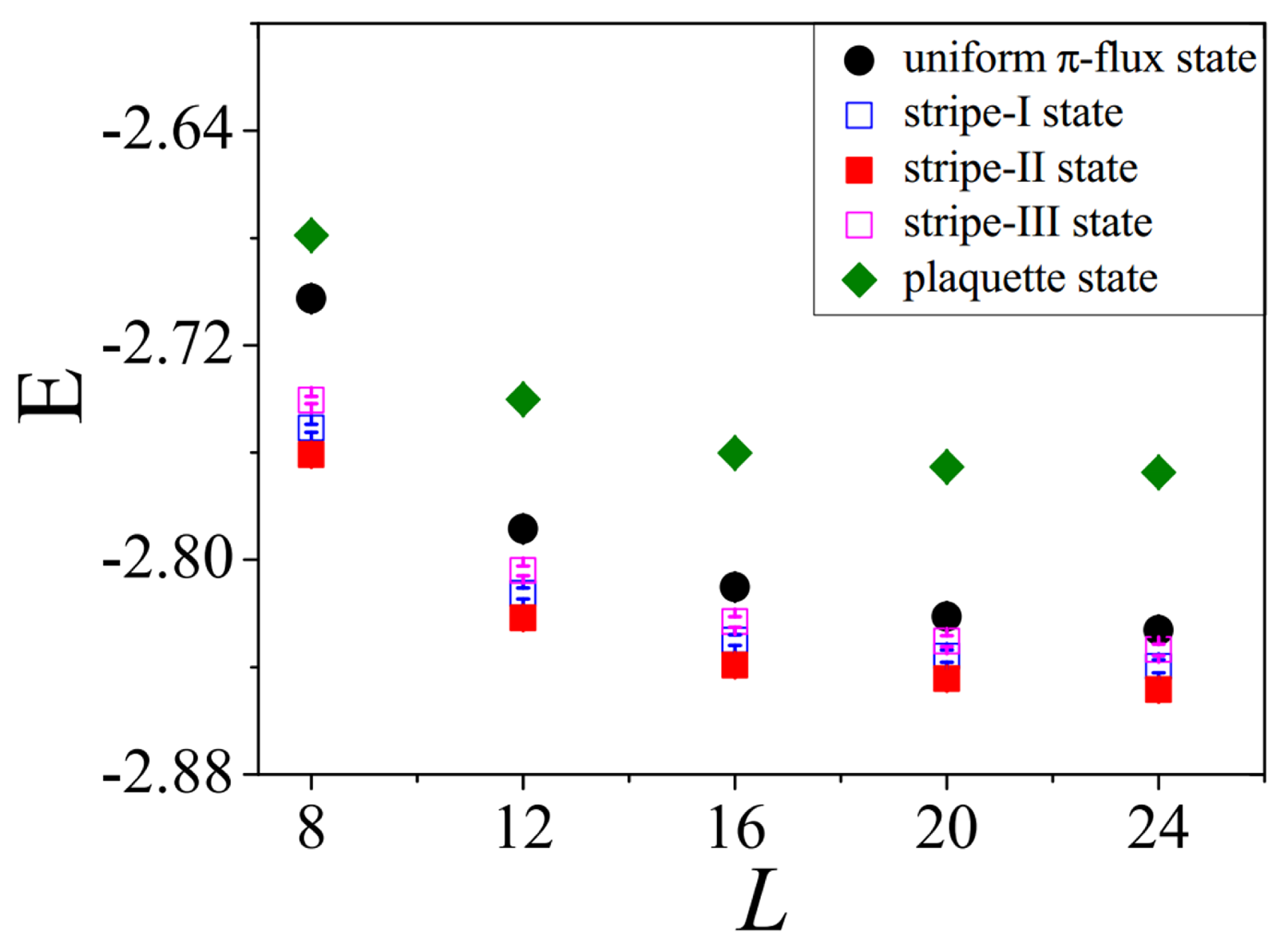}
}
\subfigure[]{
\includegraphics[width=0.48\linewidth]{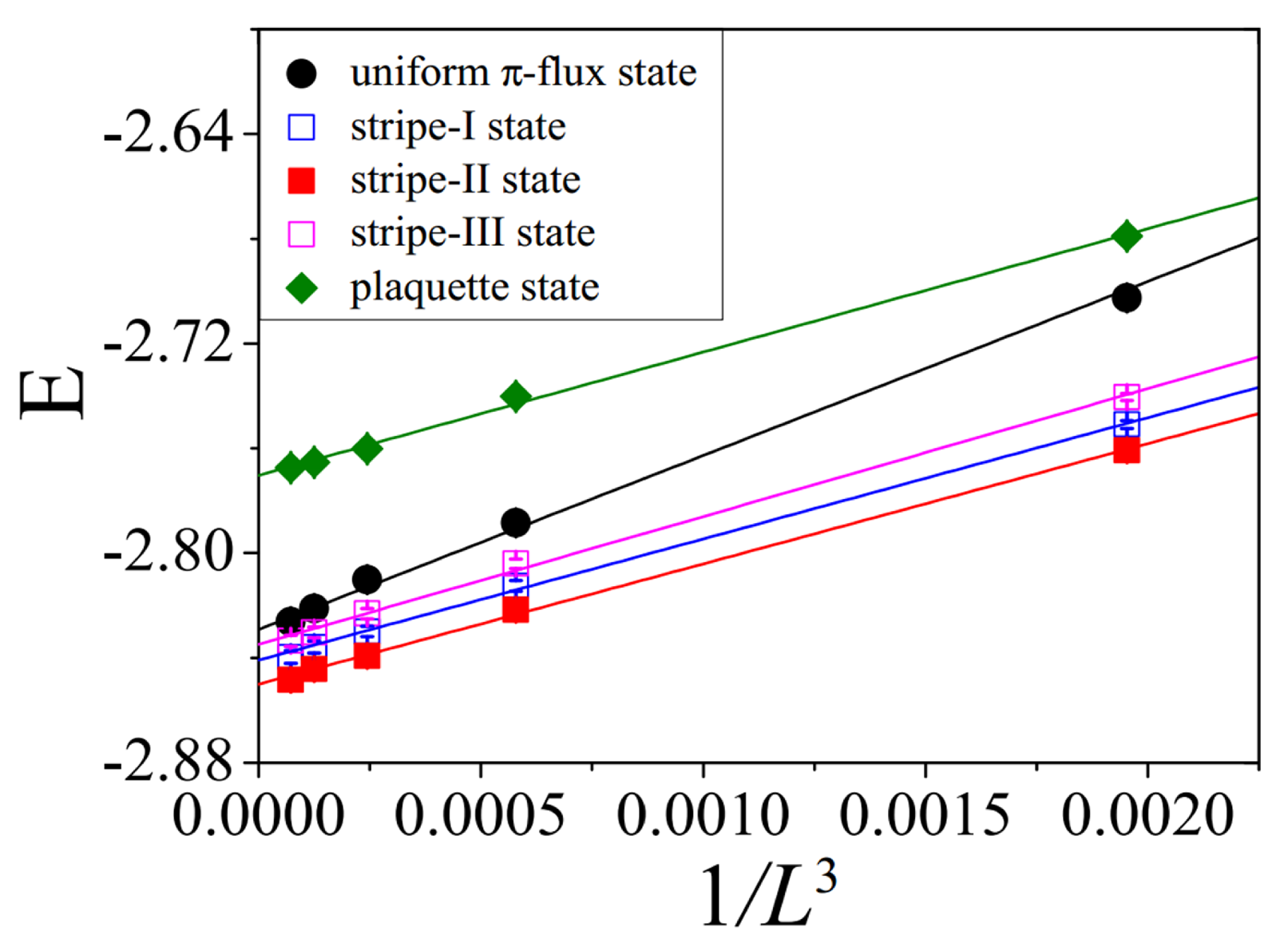}
}
\caption{All the local energy minima can be classified into the five types of states listed in Table~\ref{tab:t1}. (a) The optimized ground-state energy for the five mean-field ansatzes listed in Table~\ref{tab:t1}. The lattice geometry is chosen to be an $L\times L$ XC torus with $L=8,12,16,20,24$. (b) The linear fit of $E=E_{\infty}+\beta{}/L^3$ with $L=8,12,16,20,24$. The values of $E_{\infty}$ can be found in Table~\ref{tab:t2}.The energy error bar $(\thicksim 10^{-3})$ is approximately equal to the symbol size.} ~\label{fig:fig3}
\end{figure*}

\begin{table*}[htb]
\centering
\caption{The optimized ground-state energy and corresponding mean-field parameters for all these five types of local energy minima found in the 12 variational-parameter optimization. $E_{24}$ is the optimized ground state energy on an XC($24\times{}24$) torus. $E_{\infty}$ is obtained from the linear fit $E=E_{\infty}+\beta{}/L^3$ with $L=8,12,16,20,24$. The variational parameters $t_{n}(n=1,2,\cdots,12)$, $\delta_{1}$, $\delta_{2}$ and $\epsilon$ are defined in Eqs.~\eqref{eq:HVBC}, \eqref{eq:stripeC2}, \eqref{eq:stripyI}, \eqref{eq:stripyII}, and \eqref{eq:stripyIII}. Here we set the average $t=\frac{1}{12}\sum_{n=1}^{12}t_{n}$ as the unit of $t_n$ [see Eq.~\eqref{eq:t}].
}\label{tab:t2}
\renewcommand\arraystretch{2.5}
\setlength{\tabcolsep}{1.4ex}
\begin{tabular}{l|c|c|c|c|c}
\hline\hline
\begin{minipage}[b]{0.4\columnwidth}\raisebox{-.5\height}{\includegraphics[width=\linewidth]{p1b.pdf}}\end{minipage}
&\text{$E_{24}$}
&\text{$E_{\infty}$}
&\multicolumn{2}{c|}{$t_n$} 
& $\delta_1,\delta_2,\epsilon$\\ \hline
\text{\makecell{(a) uniform $\pi$-flux  state}}
& $-2.8263 \pm 0.0023$
& $-2.8292 \pm 0.0025$
& \multicolumn{2}{c|}{\makecell{$t_{n}|_{(n=1,\cdots,12)} \approx t;$\\ \text{[see Eq.~\eqref{eq:piflux}]}}}
& \makecell{$/$} \\ \hline

\text{(b) stripe-I state}
& $-2.8397 \pm 0.0024$
& $-2.8409 \pm 0.0023$
& \makecell{$t_1/t \approx 0.816,\,{}t_4/t \approx 1.092;$\\ \text{[see Eq.~\eqref{eq:stripyI}]}} & 
& \makecell{$\delta_1 \approx 0.145$} \\ \cline{1-4} \cline{6-6}

\text{(c) stripe-II state}
& $-2.8484 \pm 0.0026$
& $-2.8500 \pm 0.0026$
& \makecell{$t_1/t \approx 0.849,\,{}t_4/t \approx 1.151,$\\ $t_2/t\approx{}1.0;$\,\, \text{[see Eq.~\eqref{eq:stripyII}]}} & \makecell{$t_{n+6}\approx{}t_{n}|_{(n=1,\cdots,6)}$ \\ $t=(t_1+t_4)/2$}
& \makecell{$\delta_1 \approx 0.151$ \\ $\epsilon\approx{}0$} \\ \cline{1-4} \cline{6-6}

\text{(d) stripe-III state}
& $-2.8335 \pm 0.0022$
& $-2.8349 \pm 0.0028$
& \makecell{$t_1/t \approx 0.854,\,{}t_4/t \approx 1.146;$\\ \text{[see Eq.~\eqref{eq:stripyIII}]}} & 
& \makecell{$\delta_1 \approx 0.146$} \\ \hline

\text{(e) plaquette state}
& $-2.7673 \pm 0.0023$
& $-2.7703 \pm 0.0026$
& \multicolumn{2}{c|}{\makecell{$t_1/t \approx 0.58,\,{}t_4 \approx t_7 \approx {}t_2,\,{}t_{10}/t \approx 1.42,$\\ $t=(t_1+t_{10})/2;$\,\, \text{[see Eq.~\eqref{eq:plaq}]}}}
& \makecell{$\delta_1 \approx \delta_2 \approx 0.210$} \\ 
\hline\hline
\end{tabular}
\end{table*}

The main results are summarized in Fig.~\ref{fig:fig3} and Table~\ref{tab:t2}. The optimized ground state energies have been plotted as a function of the linear size $L$ [see Fig.~\ref{fig:fig3}(a)] as well as its inverse cube $1/L^3$ [see Fig.~\ref{fig:fig3}(b)]~\cite{Manousakis91,Gross1989,Neuberger1989}. Note that here the optimizations have been done with the 12 variational parameters, $t_{1,\cdots,12}$, and all the local energy minima have been classified into five types of states as listed in Table~\ref{tab:t1}. For all these five types of local energy minima found in the 12 variational parameter optimization, the optimized ground state energy and corresponding mean-field parameters can be found in Table~\ref{tab:t2}.

Using the formula $E(L)=E_{\infty}+\beta/L^3$~\cite{Manousakis91,Gross1989,Neuberger1989}, extrapolation to the thermodynamic limit $L\to{}\infty$ allows us to draw the third conclusion:
\begin{theorem} 
Among all the five types of ansatzes listed in Table~\ref{tab:t1}, the stripe-II state has the lowest ground-state energy.
\end{theorem}

{\bf Stability of variational ground states:} We will now investigate the energetic stability of these variational ground states. To do this, we introduce a Gaussian noise for a given mean-field ansatz $\{t_{i}\}$ such that $t_{n}\rightarrow{}t_{n}+\delta{}t_{n}$, where ${\delta{}t_n}\sim{}0.01{}t_n$ and the constraint $\sum_{n}\delta{}t_n=0$ has been imposed. We then initialize the VMC computations with these noisy ansatzes and allow all 12 parameters $t_{n=1,\cdots,12}$ to be optimized during the VMC computations. 

We have observed that the stripe-II state and the plaquette state are stable against noise in VMC simulations, even when a relatively strong Gaussian noise $\{\delta{}t_i\}$ is added. This means that a noisy stripe-II state converges to a stripe-II state after VMC optimization, and so does a noisy plaquette state. In contrast, the uniform $\pi$-flux state, the stripe-I state, and the stripe-III state are inherently unstable to Gaussian noise. They easily converge to a stripe-II state instead of the original state type during VMC optimization.

\begin{figure}[tb]
\centering
\includegraphics[width=\linewidth]{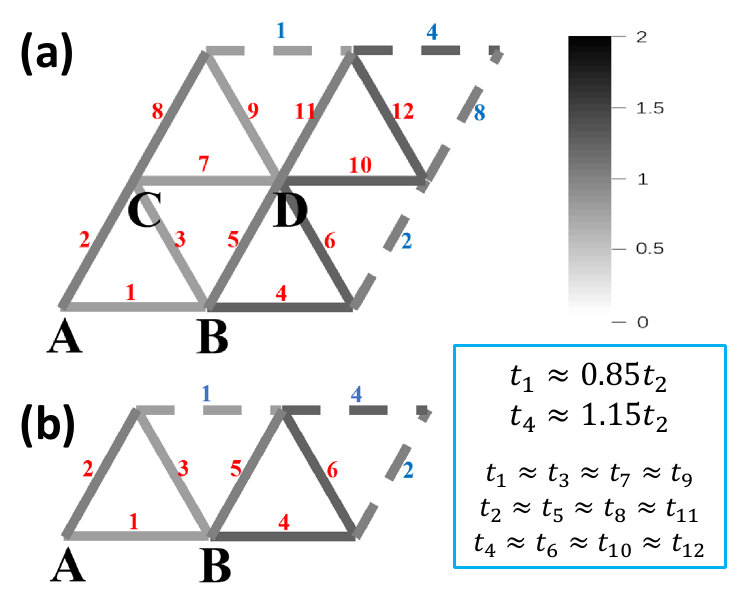}
\caption{Stipe-II state on an XC($24 \times 24$) torus. The lowest energy state is characterized by 12 variational parameters $t_{n=1,\cdots,12}$, which are illustrated in a gray color-map. (a) A large unit cell on the XC($24 \times 24$) torus for a $2\times{}2$ generic VBC state. The dashed lines indicate PBCs. (b) A unit cell for the strip-II state. Here $t_1+t_4=2t_2$ or $\epsilon=0$ [defined in Eq.~\eqref{eq:stripyII}] is found.} ~\label{fig:stripe-II}
\end{figure}

{\bf Most stable ground state:} We now discuss the most stable ground state, the stripe-II state defined in eq.~\eqref{eq:stripyII}. First, we illustrate in Fig.~\ref{fig:stripe-II} the set of all 12 variational parameters $\{t_1,t_2,\cdots,t_{12}\}$ for the fully optimized state on an XC($24\times{}24$) torus. Note that such a state satisfies the constraints $t_{n+6}=t_{n}|_{n=1,\cdots,6}$ and $t_2=t_5$ within the error bar, which defines a stripe-II state. Moreover, this stripe-II state reaches its lowest energy at $t_1+t_4=2t_2$ or $\epsilon=0$ in eq.~\eqref{eq:stripyII}. 

\begin{figure}[tb]
\centering
\includegraphics[width=\linewidth]{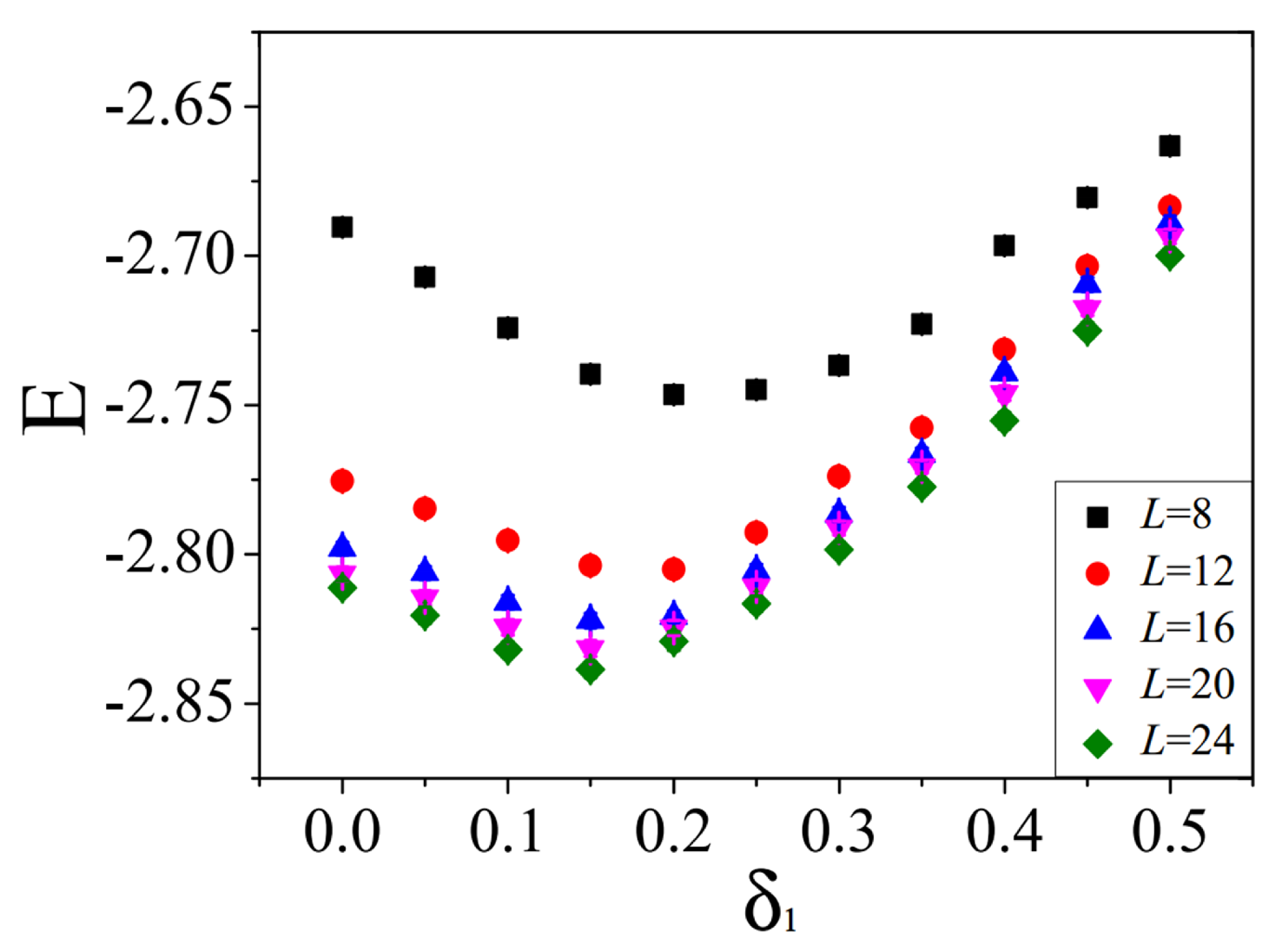}
\caption{Stipe-II states on an XC($L\times{}L$) torus. Here we have set $\epsilon=0$ in Eq.~\eqref{eq:stripyII}. The ground state energy $E(\delta_1,\epsilon=0)$ is plotted as a function of $\delta_1$ for $L=8,12,16,20,24$. The energy error bar $(\thicksim 10^{-3})$ is approximately equal to the symbol size. The energy minimum is found to be $E_{min}=-2.8386\pm{}0.0027$ with $\delta_1=0.15$ on the $L=24$ lattice.} ~\label{fig:stripe-II-ansatz}
\end{figure}

Second, we will focus on the stripe-II state defined in Eq.~\eqref{eq:stripyII}, which is characterized by two independent parameters $\delta_1$ and $\epsilon$. Motivated by the above result of the stripe-II state during the 12-parameter optimization, we fix $\epsilon=0$ and explore the ground state energy by varying $\delta_1$ in Eq.~\eqref{eq:stripyII}.
The ground state energy $E(\delta_1, \epsilon=0)$ was calculated on a $L\times{}L$ torus and plotted in Fig.~\ref{fig:stripe-II-ansatz} as a function of $\delta_1$, considering different system sizes of $L=8, 12, 16, 20, 24$. Remarkably, the minimum energy was found to be $E_{min}=-2.8386\pm{}0.0027$ with $\delta_1=0.15$ on a $L=24$ lattice, showing a strong agreement with the results obtained by the 12-parameter optimization.

\begin{figure*}[htb]
\centering
\includegraphics[width=0.8\linewidth]{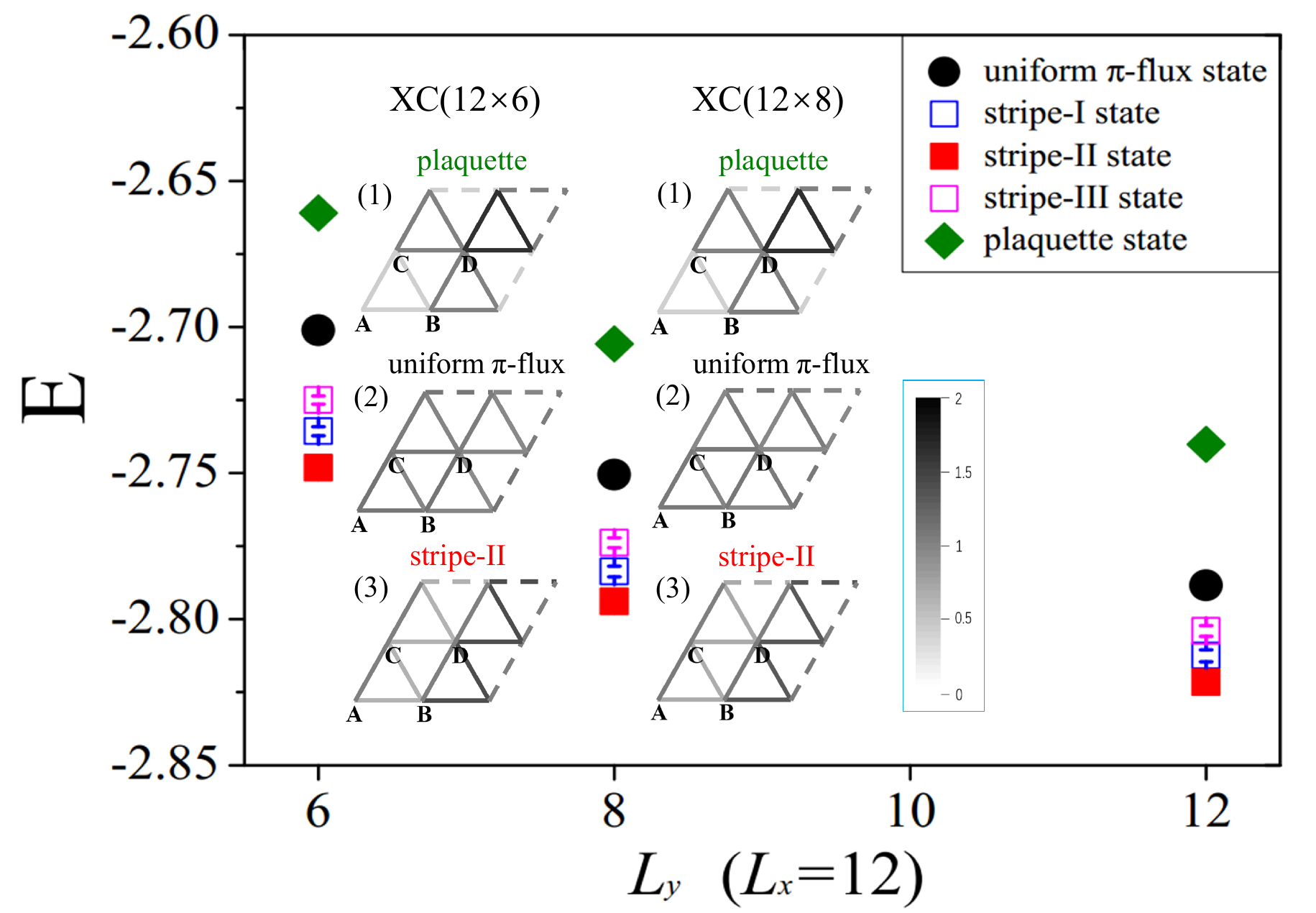}
\caption{The optimized energy $E(t_1,\cdots,t_{12})$ for different shapes of XC($L_x\times{}L_y$) tori. Here $L_x=12$ is fixed and $L_y$ is chosen to be $L_y=6,8,12$. Inserts (1), (2), and (3) represent three types of states in a gray color map. The energy error bar $(\thicksim 10^{-3})$ is approximately equal to the symbol size.} ~\label{fig:size}
\end{figure*}

{\bf{}Finite-size effect:} Finally, we would like to examine the finite-size effect on lattices with different aspect ratios, $L_{x}/L_{y}$. To do this, we fix $L_x$ and vary $L_y$, and then perform VMC calculations on the XC($L_x\times{}L_y$) tori by the 12-parameter optimization. A number of typical results are shown in Fig.~\ref{fig:size}. We find that the stripe-II state remains the lowest energy state as the aspect ratio $L_{x}/L_{y}$ varies, i.e. our conclusions remain unaffected by variations in the aspect ratio.

\section{Summary}\label{sec:sec5}

In summary, the VMC method has been exploited to study the SU(4) symmetric spin-orbital model, i.e., the so-called Kugel-Khomskii model, on the triangular lattice. Begin with a generic $2\times{}2$ valence-bond-crystal state that is characterized by 12 variational parameters $\{t_1,t_2,\cdots,t_{12}\}$, we have performed extensive VMC calculations and found that (i) all the local energy minima exhibit a $\pi$-flux configuration, (ii) the optimized local energy minimum must belong to one of five types of states that include the uniform $\pi$-flux state, three $C_2$- stripe states (dubbed stripe-I, stripe-II, and stripe-III states), and the plaquette state, as listed in Table~\ref{tab:t1}, and (ii) the stripe-II state has the lowest ground state energy with an extrapolated value of $E_{\infty}=-2.8500 \pm 0.0026$ in the thermodynamic limit.

This strip-II state breaks the lattice $C_6$ rotational symmetry to a $C_2$ rotational symmetry and has $2\times{}1$ unit cells. Since the SU(4) spin-orbital rotational symmetry is respected, such a state is indeed a nematic quantum spin-orbital liquid state. The low energy excitations on top of the ground state are gapless and can be represented by a parton Fermi surface consisting of open orbits in the Brillouin zone. All of these results are in excellent agreement with previous results obtained from the Gutzwiller-boosted DMRG~\cite{jin2022Bulletin}.

\begin{acknowledgments}
We would like to thank Rong-Yang Sun and Hong-Hao Tu for their helpful discussions. This work is supported in part by National Key Research and Development Program of China (No. 2022YFA1403403),
National Natural Science Foundation of China (No.12274441, 12034004), and the European Research Council (ERC) under the European Unions Horizon 2020 research and innovation program (Grant agreement No.~771537).
\end{acknowledgments}

\appendix

\setcounter{table}{0}   
\setcounter{figure}{0}
\renewcommand{\thetable}{A\arabic{table}}
\renewcommand{\thefigure}{A\arabic{figure}}

\section{Stochastic reconfiguration method in VMC calculations}

To perform the variational Monte Carlo (VMC) optimization, a stochastic reconfiguration (SR) method has been adopted. It simulates the effect of the Hessian matrix with a more cost-effective approach, which we now review briefly. 

When carrying out the VMC optimization, one key step is the computation of the expectation value of Hamiltonian and its gradient with respect to the variational parameters. Without loss of generality, the variational state $|\Psi\rangle$ can be constructed by the wave function $|\Psi(C)\rangle$ in an orthonormal basis $|C\rangle$, written as
\setcounter{equation}{0}
\renewcommand\theequation{A\arabic{equation}}
\begin{equation}
|\Psi\rangle=\sum_C \Psi(C)|C\rangle.  \label{eq:A1}
\end{equation}
Then, the variational energy is given by
\begin{equation}
E=\frac{\langle \Psi |H| \Psi \rangle}{\langle \Psi \mid \Psi \rangle}=\frac{\sum_C |\Psi(C)|^2 E_{loc}(C)}{\sum_C |\Psi(C)|^2}.  \label{eq:A2}
\end{equation}
The local energy $E_{loc}$ is sampled and its average value is calculated by:
\begin{equation}
E_{loc}(C)=\sum_{C'}\frac{\langle C|H|C' \rangle \Psi(C')}{\Psi(C)}.  \label{eq:A3}
\end{equation}

To search for the minimum energy of $E(t_1,\cdots,t_{12})$ in which a value taking condition $\sum_{n=1}^{12}t_n=12$ has been imposed. For convenience, the series of parameters in wave function $|\Psi(t_1,t_2,\cdots,t_{12})\rangle$ have been abbreviated as $\{\theta\}$. The gradient of the variational energy with respect to the variational parameters $\{\theta\}$ is given by 
\begin{equation}
\begin{split}
 \nabla_{\theta}E(\theta)= & 2\langle\nabla_{\theta}(ln\Psi(C))E_{loc}(C)\rangle \\
 & -2\langle E_{loc}(C)\rangle \langle \nabla_{\theta}(ln \Psi(C))\rangle, \label{eq:A4}
\end{split}
\end{equation}
which is obtained by the derivative with respect to $\{\theta\}$ of the following formula 
\begin{equation}
E(\theta)=\frac{\langle \Psi(\theta)|H|\Psi(\theta) \rangle}{\langle \Psi(\theta)|\Psi(\theta) \rangle}.
\label{eq:A5}
\end{equation}
In the formula \eqref{eq:A4}, the two quantities can be calculated by standard Monte Carlo sampling on the distribution generated by $|\Psi(\theta)|^2$, and the quantity $\langle \nabla_{\theta}(ln \Psi(C))\rangle$ will play a key role in the calculation of the gradient.

To obtain the adjustment direction and the step length of the $\{\theta\}$, an approximate Hessian matrix $\langle \Psi(\theta)|H|\dot{\Psi}(\theta) \rangle$ requires the construction
\begin{equation}
\frac{\partial^2}{\partial \theta \partial \theta'} \frac{\langle \Psi(\theta)|\Psi(\theta') \rangle}{\langle \Psi(\theta)|\Psi(\theta) \rangle}. \label{eq:A6}
\end{equation}
As the essence of the SR method, the second derivative of energy with respect to parameters $\{\theta\}$ can be obtained by matrix~\eqref{eq:A6}. For imitating the effect of the Hessian matrix, a positive-definite Hermitian matrix $S$ generated from the metric of the variational state in the variational space is proposed. More specifically, the matrix element is given by
\begin{equation}
\begin{split}
S_{ij}= & \langle \nabla_{\theta}[ln \Psi(C)]_i \nabla_{\theta}[ln \Psi(C)]_j \rangle -\langle \nabla_{\theta}[ln \Psi(C)]_i\rangle \langle \nabla_{\theta}[ln \Psi(C)]_j\rangle, \label{eq:A7}
\end{split}
\end{equation}
where the index $i$ is with respect to the $i$-th variational paramter.

Thanks to the quantity $\langle\nabla_{\theta}(ln\Psi(C)) \rangle$ derived  from Eq.\eqref{eq:A4} above, an approximate Hessian matrix $S$ can be obtained from Eq.~\eqref{eq:A7}, 
Therefore, the direction and step length for performing parameters optimization can be found as follows,
\begin{equation}
\Delta\vec{\theta}=-\frac{1}{2\kappa}S^{-1}\vec{g}. \label{eq:A8} 
\end{equation}
Here $\kappa$ is an empirical positive parameter, which is adjusted manually and defines the step length of parameter optimization.

\section{VMC calculations for four Gutzwiller projected states listed in Table~\ref{tab:t1}}

\begin{table*}[tb]
\centering
\caption{The optimized ground states for four types of Gutzwiller projected states listed in Table~\ref{tab:t1}. Here $E_L$ is the optimized ground state energy per site on an XC($L \times{}L$) torus with $L=12,16,24$. The variational parameters  $\delta_{1}$, $\delta_{2}$ and $\epsilon$ are defined in Eqs.~\eqref{eq:HVBC}, \eqref{eq:stripeC2}, \eqref{eq:stripyI}, \eqref{eq:stripyII}, and \eqref{eq:stripyIII}, respectively.
}\label{tab:ta1}
\renewcommand\arraystretch{2.5}
\setlength{\tabcolsep}{1.5ex}
\begin{tabular}{l|c|c|c|c|c|c}
\hline\hline
&\multicolumn{2}{c|}{$L=12$}
&\multicolumn{2}{c|}{$L=16$}
&\multicolumn{2}{c}{$L=24$} \\ \cline{2-7}

&\text{$E_{L}$}
& $\delta_1,\delta_2,\epsilon$
&\text{$E_{L}$}
& $\delta_1,\delta_2,\epsilon$
&\text{$E_{L}$}
& $\delta_1,\delta_2,\epsilon$ \\ \hline

\text{\makecell{(a) stripe-I state}}
& $ -2.8091 \pm 0.0014$
& \makecell{$\delta_1 \approx 0.174$}
& $ -2.8276 \pm 0.0021$
& \makecell{$\delta_1 \approx 0.156$}
& $ -2.8377 \pm 0.0024$
& \makecell{$\delta_1 \approx 0.145$} \\ \hline

\text{(b) stripe-II state}
& $ -2.8174 \pm 0.0015$
& \makecell{$\delta_1 \approx 0.181$ \\ $\epsilon\approx{}0$}
& $ -2.8369 \pm 0.0021$
& \makecell{$\delta_1 \approx 0.161$ \\ $\epsilon\approx{}0$}
& $ -2.8462 \pm 0.0024$
& \makecell{$\delta_1 \approx 0.151$ \\ $\epsilon\approx{}0$} \\ \hline

\text{(c) stripe-III state}
& $ -2.8015 \pm 0.0014$
& \makecell{$\delta_1 \approx 0.177$}
& $ -2.8210 \pm 0.0020$
& \makecell{$\delta_1 \approx 0.156$}
& $ -2.8312 \pm 0.0025$
& \makecell{$\delta_1 \approx 0.145$} \\ \hline

\text{(d) plaquette state}
& $ -2.7401 \pm 0.0010 $
& \makecell{$\delta_1 \approx 0.218$ \\ $\delta_2 \approx 0.217$}
& $ -2.7600 \pm 0.0018 $
& \makecell{$\delta_1 \approx \delta_2 \approx 0.203$}
& $ -2.7670 \pm 0.0020 $
& \makecell{$\delta_1 \approx \delta_2 \approx 0.208$} \\ 
\hline\hline

\end{tabular}
\end{table*}

In the main text, using the SR method, we have searched for local energy minima on XC($L\times{L}$) tori up to $L=24$ (see Fig.~\ref{fig:fig3} and Table~\ref{tab:t2}) with the generic 12-parameter $2\times{}2$ VBC states given in Eq.~\eqref{eq:HVBC}.
In this appendix, we perform standard VMC calculations and search for local energy minima for four Gutzwiller projected states on XC($L\times{}L$) tori with $L=12,16,24$. These Gutzwiller projected states are characterized by the one or two variational parameters, as listed in Table~\ref{tab:t1}, namely, $\delta_1$ in Eq.~\eqref{eq:stripyI} and Eq.~\eqref{eq:stripyIII} [stripe-I and stripe-III states], $\delta_1$ and $\epsilon$ in Eq.~\eqref{eq:stripyII} [stripe-II state], and $\delta_1$ and $\delta_2$ in Eq.~\eqref{eq:plaq} [plaquette state].

The optimized ground state energy and corresponding variational parameters are given in Table~\ref{tab:ta1}, which are consistent with the 12-parameter optimization results shown in Table~\ref{tab:t2} and Fig.~\ref{fig:fig3} in the main text.

\bibliography{paper1.bib}

\end{document}